\documentclass[twocolumn,showpacs,aps,prl,amsmath,amssymb,nobibnotes,floatfix]{revtex4-1}
\newcommand{\beq}{\begin{equation}}
\newcommand{\eeq}{\end{equation}}
\newcommand{\beqn}{\begin{eqnarray}}
\newcommand{\eeqn}{\end{eqnarray}}

\usepackage{graphicx}
\usepackage{epsf}

\usepackage{graphics,epsfig,placeins,subfigure,wrapfig,amssymb,pifont}

\usepackage{psfrag}
\usepackage[usenames]{color}
\usepackage{multirow}
\usepackage{rotating}
\usepackage{hyperref}
\usepackage{ragged2e}

\begin{document}
\title{One-arm Spiral Instability in Hypermassive Neutron Stars
  \\ Formed by Dynamical-Capture Binary Neutron Star Mergers}

\author{Vasileios Paschalidis${}^1$, William E.\ East${}^2$, Frans
  Pretorius${}^{1}$, and Stuart L. Shapiro${}^{3,4}$} 
\affiliation{
  ${}^1$Department of Physics, Princeton University, Princeton, NJ 08544, USA  \\
  ${}^2$Kavli Institute for Particle Astrophysics and Cosmology, Stanford University, SLAC
  National Accelerator Laboratory, Menlo Park, CA 94025, USA\\
${}^3$Department of Physics, University of Illinois at Urbana-Champaign, Urbana, IL 61801\\
${}^4$Department of Astronomy \& NCSA, University of Illinois at Urbana-Champaign, Urbana, IL 6180
}

\begin{abstract}
Using general-relativistic hydrodynamical simulations, we show that merging
binary neutron stars can form hypermassive neutrons stars that undergo the
one-arm spiral instability. We study the particular case of a dynamical
capture merger where the stars have a small spin, as may arise in globular
clusters, and focus on an equal-mass scenario where the spins are aligned
with the orbital angular momentum.
We find that this instability develops when post-merger fluid vortices
lead to the generation of a toroidal remnant --- a configuration whose
maximum density occurs in a ring around the center-of-mass --- with
high vorticity along its rotation axis.  The instability quickly
saturates on a timescale of $\sim 10$ ms, with the $m=1$ azimuthal
density multipole mode dominating over higher modes.
The instability also leaves a characteristic imprint on the
post-merger gravitational wave signal that could be detectable if the
instability persists in long-lived remnants.

\end{abstract}

\maketitle


{\em Introduction.}---%
The possibility of observing the inspiral and merger of neutron
star--neutron star (NSNS) binaries is an exciting prospect that soon
may be realized.  Often referred to as ``multimessenger'' sources,
NSNSs emit copious amounts of gravitational waves (GWs), and hence are
a primary target of ground-based laser interferometers such as
aLIGO~\cite{LIGO}, VIRGO~\cite{VIRGO} and KAGRA~\cite{Somiya:2011me},
and may generate transient electromagnetic (EM) signals, both
before~\cite{Hansen:2000am,McWilliams:2011zi,Paschalidis:2013jsa,PalenzuelaLehner2013,2014PhRvD..90d4007P}
and after~\cite{MetzgerBerger2012,2015MNRAS.446.1115M} merger. These
EM transients could be observed by current and future telescopes, such
as PTF~\cite{2009PASP..121.1334R}, PanSTARRS~\cite{PanSTARRS}, and
LSST~\cite{2012arXiv1211.0310L}. By combining GW and EM signals from
NSNSs one can in principle test relativistic gravity and constrain the
behavior of matter at super-nuclear densities.  Furthermore, NSNS
mergers may offer explanations to long-standing astrophysical puzzles,
such as the nature of short-hard gamma ray burst
progenitors~\cite{PiranSGRB_review,Meszaros:2006rc,PaschalidisJet2015},
and the origin of r-process elements~\cite{Rosswog:1998gc}.

The interpretation of EM and GW signals from NSNS mergers will rely on
a solid theoretical understanding of these events. Such understanding
requires simulations in full general relativity (GR) to treat both the
rapidly varying, strong field spacetime and the relativistic
velocities that naturally arise in these events. There have been
numerous such studies, mostly focusing on quasicircular inspiral and
mergers (see, e.g.~\cite{faber_review} for a review,
and~\cite{Paschalidis2012,Neilsen2014,
  Dionysopoulou2015,Sekiguchi2015,Dietrich2015,Palenzuela2015} for
recent work), but also some simulations of eccentric inspiral and
mergers~\cite{gold,East2012NSNS}.  The latter binaries may be
dynamically assembled in dense stellar systems such as globular
clusters (GCs) through single-single~\cite{Kocsis:2011jy,lee2010} or
binary-single star interactions~\cite{Samsing:2013kua}. Although the
rates are very uncertain, they have been estimated to be as high as
$\sim50\rm{\ yr}^{-1}\ Gpc^{-3}$~\cite{lee2010}. Note though that the
majority of events sourced by binary-single interactions will enter
the aLIGO frequency band ($\gtrsim10$ Hz) as low eccentricity systems.
Also, though a recent study of dynamically assembled hierarchical
triple systems in GCs found Lidov-Kozai induced merger of the inner
binary could lead optimistically to several aLIGO detections per year
of highly eccentric {\em black hole} binaries, they estimated that
this channel would offer a negligible contribution to NSNS merger
rates~\cite{Antonini2015}.
Another aspect of NSNS systems dynamically assembled in GCs important
to the work presented here is that (regardless of eccentricity at
merger) the individual NSs are likely to have non-negligible spin,
given the large population of millisecond pulsars (MSPs) found there
(see~\cite{EPP2015} for further discussion on the relevance of NS spin
in compact binaries).

A NSNS merger may not immediately form a black hole (BH), but instead
result in a hypermassive NS (HMNS)---a long-lived, but transient
remnant that is supported against collapse by differential rotation
and thermal energy.  Here we report results from a simulation where a
HMNS forms after the eccentric merger of two equal mass NSs that each
have a spin
period of 10.6 ms. An important feature we discover is that the HMNS
develops the so-called one-arm ($m=1$) spiral instability. This
instability grows from seeds at the level of numerical truncation
error to dominate eventually the azimuthal perturbations of the star.
In a follow-up work~\cite{followup} we will present results from a
broader range of initial conditions, in particular asymmetric cases
where the initial data does contain a small $m=1$ component,
suggesting that our results are robust and not at artifact of
truncation error.  Since the qualitative features of the instability
do not seem to depend on how we seed the initial mode, 
here we restrict discussion to this one case.

The one-arm instability was first seen in Newtonian simulations of
differentially rotating stars with soft equations of state
(EOSs)~\cite{Centrella2001}, and argued to be triggered by a toroidal
structure in the stellar density profile~\cite{Saijo2003}.  Motivated
by~\cite{Watts2005}, \cite{Saijo2006} suggested that this instability
develops near the corotation radius of the HMNS, i.e. where the
azimuthal pattern speed of the unstable mode is commensurate with the
local angular velocity of fluid elements of the star.
Newtonian~\cite{Ou2006} and general-relativistic~\cite{Corvino2010}
simulations of isolated rotating stars seem to confirm this
interpretation. The one-arm spiral instability can develop in isolated
stars even for stiff equations of state~\cite{Ou2006},
and has been found to occur in proto-NSs formed in
Newtonian~\cite{Ott2005} and
general-relativistic~\cite{Ott2007PhRvL,Kuroda2014} core-collapse
simulations.  Although it has been speculated that it could operate in
the HMNS remnants of NSNS mergers~\cite{Saijo2003}, the one-arm spiral
instability has not been reported to occur in NSNS mergers until now.
Here we demonstrate using GR hydrodynamic simulations that the
instability can develop in a NSNS dynamical capture merger remnant,
and offer a description of how the process unfolds in terms of
post-merger vortex dynamics. We also show that the mode produces a
strong $m=1$ component to the GW signal, which, if sufficiently
long-lived, could be observable by aLIGO.

\begin{figure} 
\begin{center} 
\includegraphics[height = 1.66in,angle=-90]{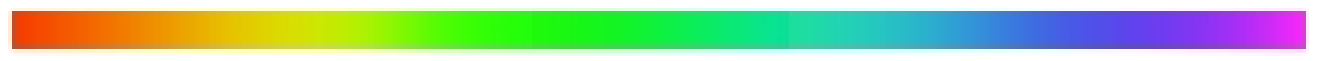}
\put(-17,2.5){$10^{15}$}
\put(-120,2.5){$10^{12}$gm/cm$^{3}$}
\hspace{0.005 cm}
\includegraphics[height = 1.66in,angle=-90]{vertical_scale.eps}
\put(-16,2.5){$150$}
\put(-120,2.5){$7.5$rad/ms}
\hspace{0.005 in}
\includegraphics[trim =5.5cm 2.20cm 5.5cm 2.20cm,height=1.4in,clip=true,draft=false]{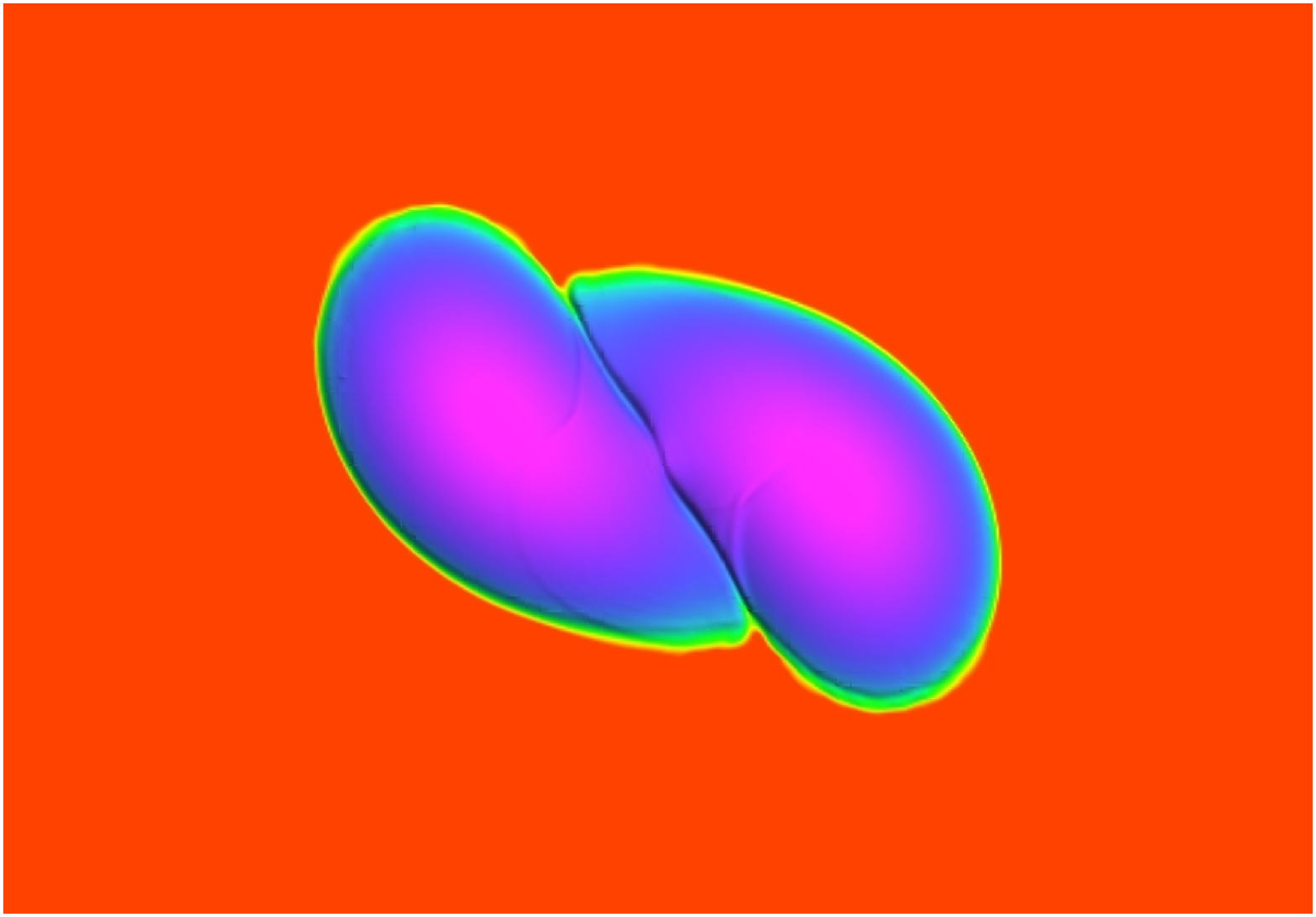}
\includegraphics[trim =5.5cm 2.20cm 5.5cm 2.20cm,height=1.4in,clip=true,draft=false]{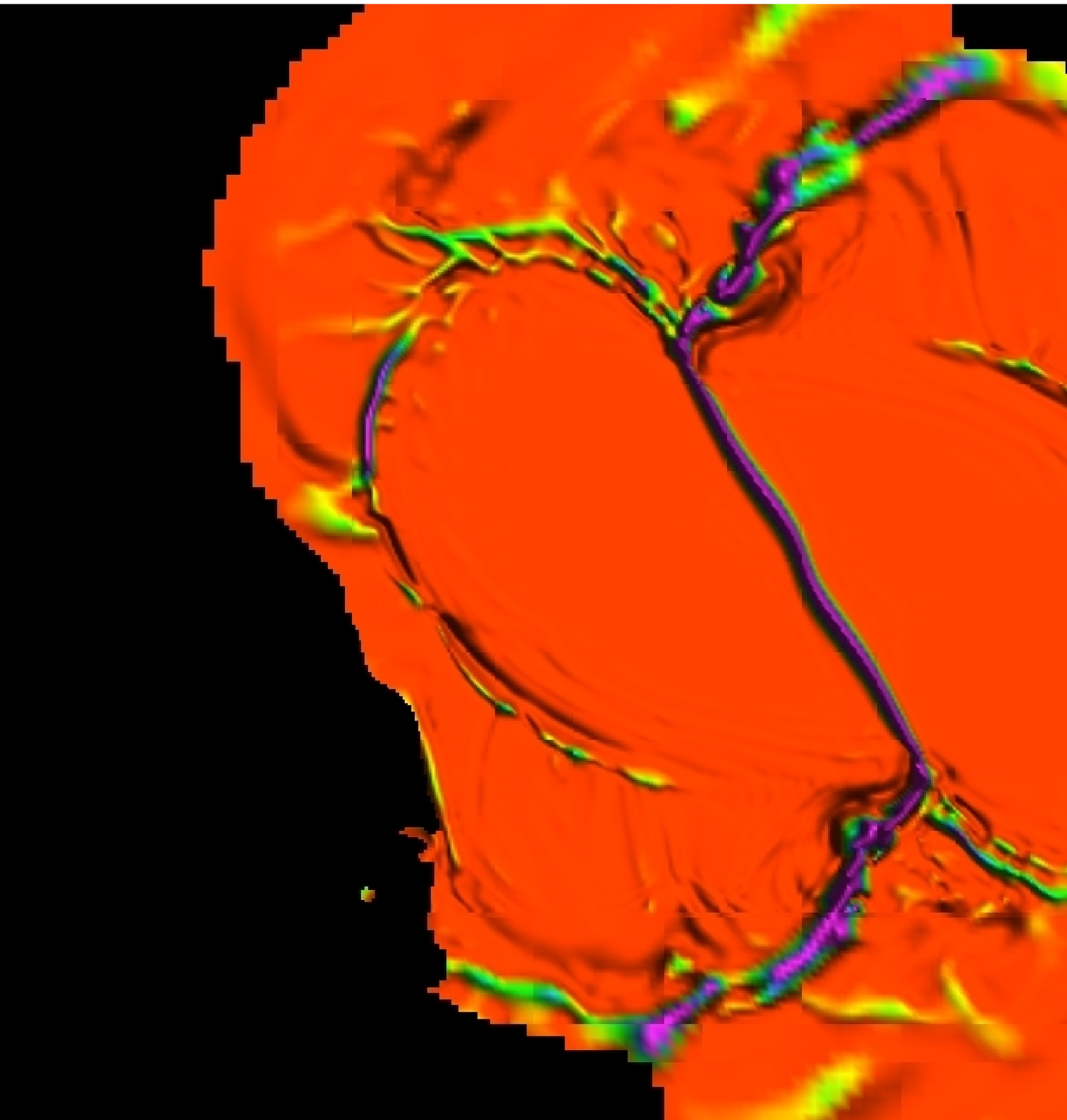}
\includegraphics[trim =5.5cm 2.20cm 5.5cm 2.20cm,height=1.4in,clip=true,draft=false]{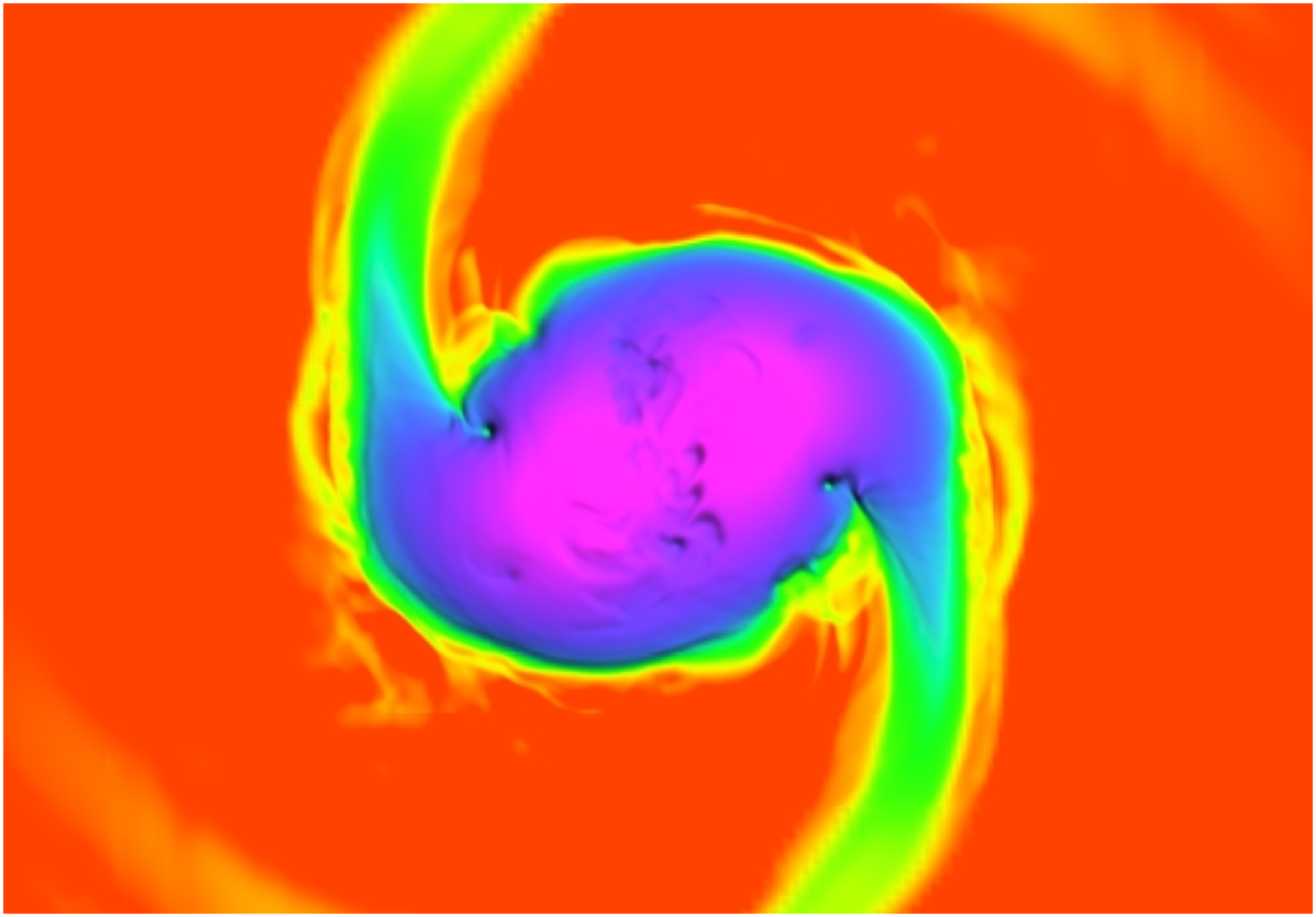}
\includegraphics[trim =5.5cm 2.20cm 5.5cm 2.20cm,height=1.4in,clip=true,draft=false]{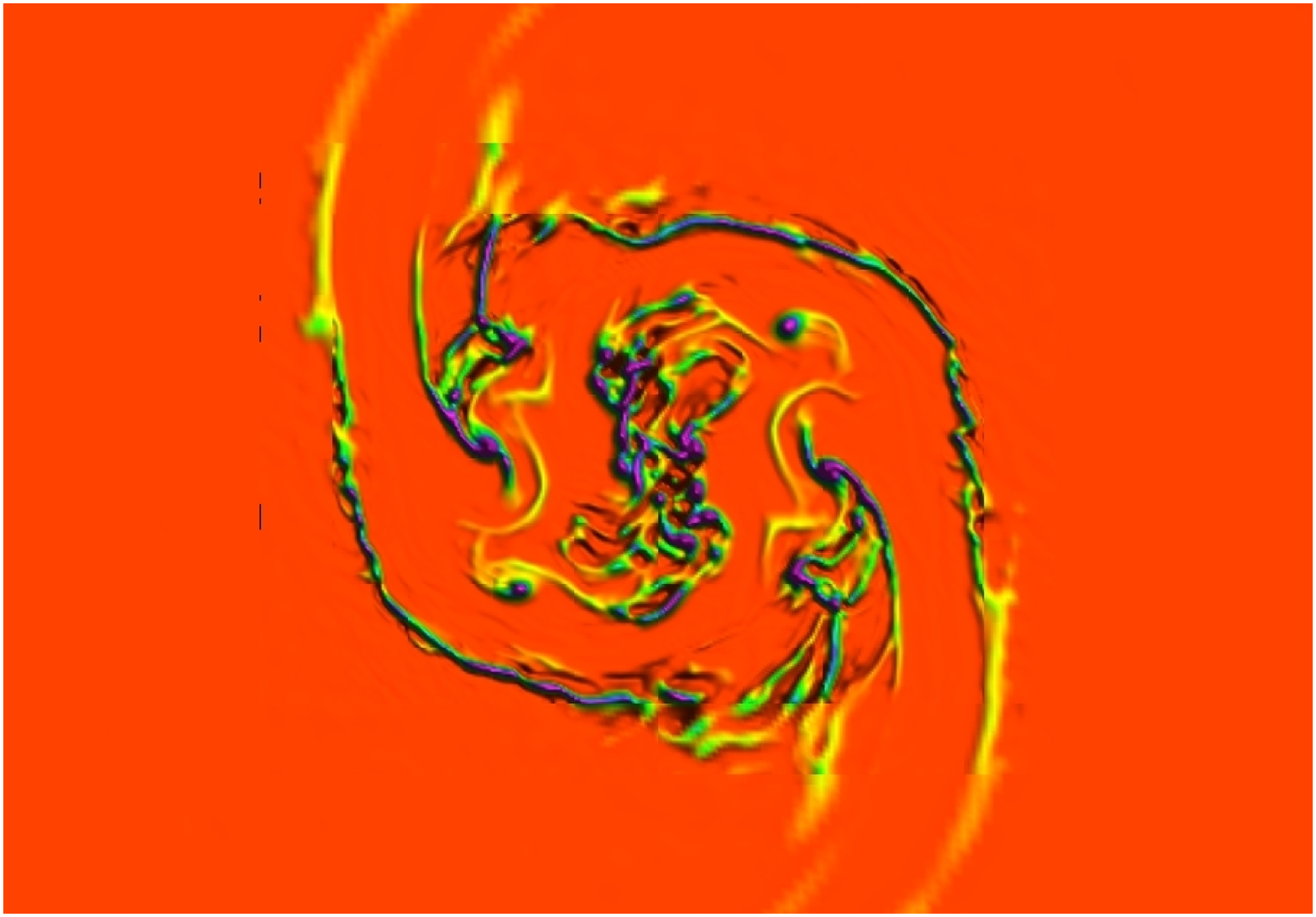}
\includegraphics[trim =5.5cm 2.20cm 5.5cm 2.20cm,height=1.4in,clip=true,draft=false]{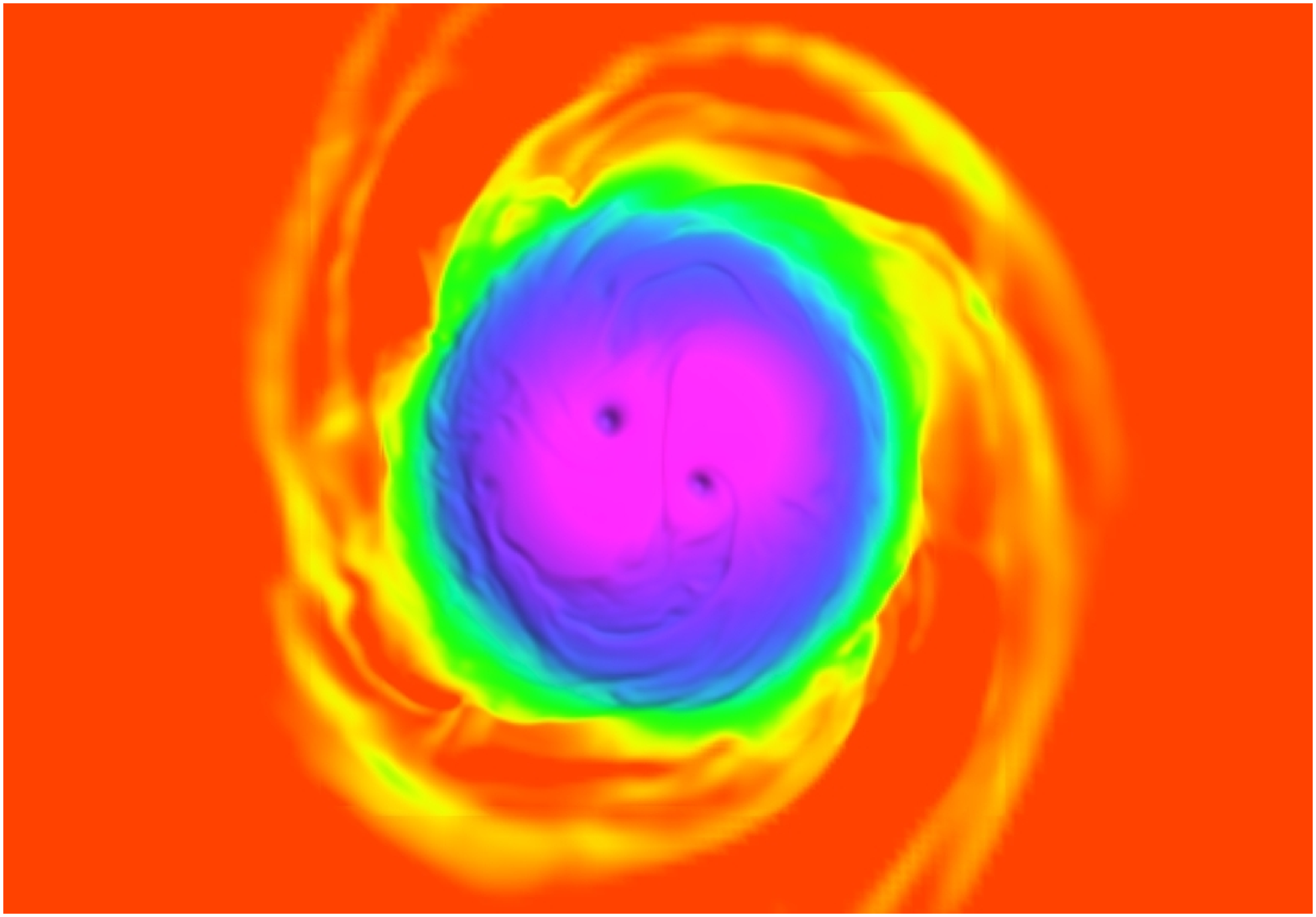}
\includegraphics[trim =5.5cm 2.20cm 5.5cm 2.20cm,height=1.4in,clip=true,draft=false]{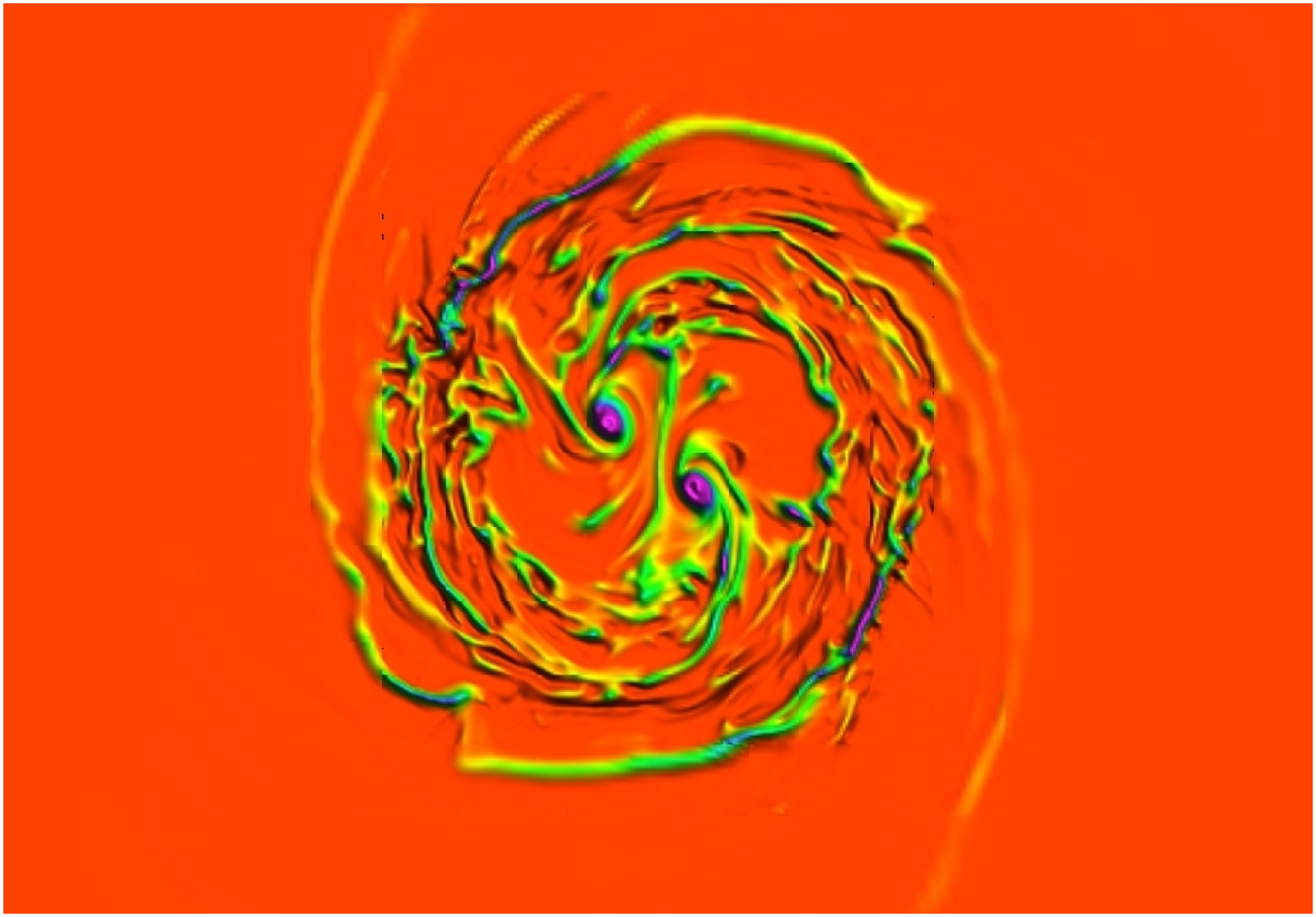}
\includegraphics[trim =5.5cm 2.20cm 5.5cm 2.20cm,height=1.4in,clip=true,draft=false]{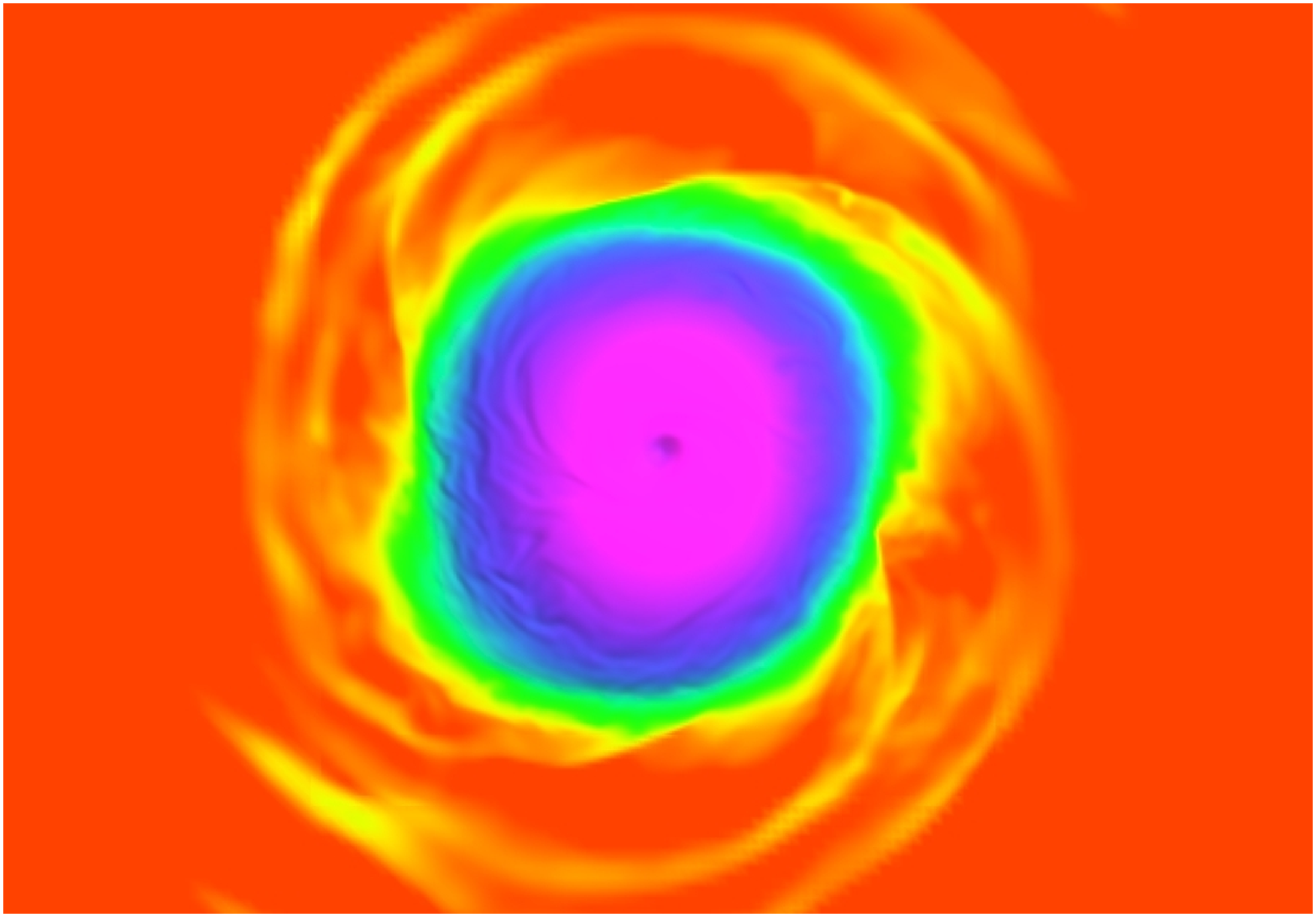}
\includegraphics[trim =5.5cm 2.20cm 5.5cm 2.20cm,height=1.4in,clip=true,draft=false]{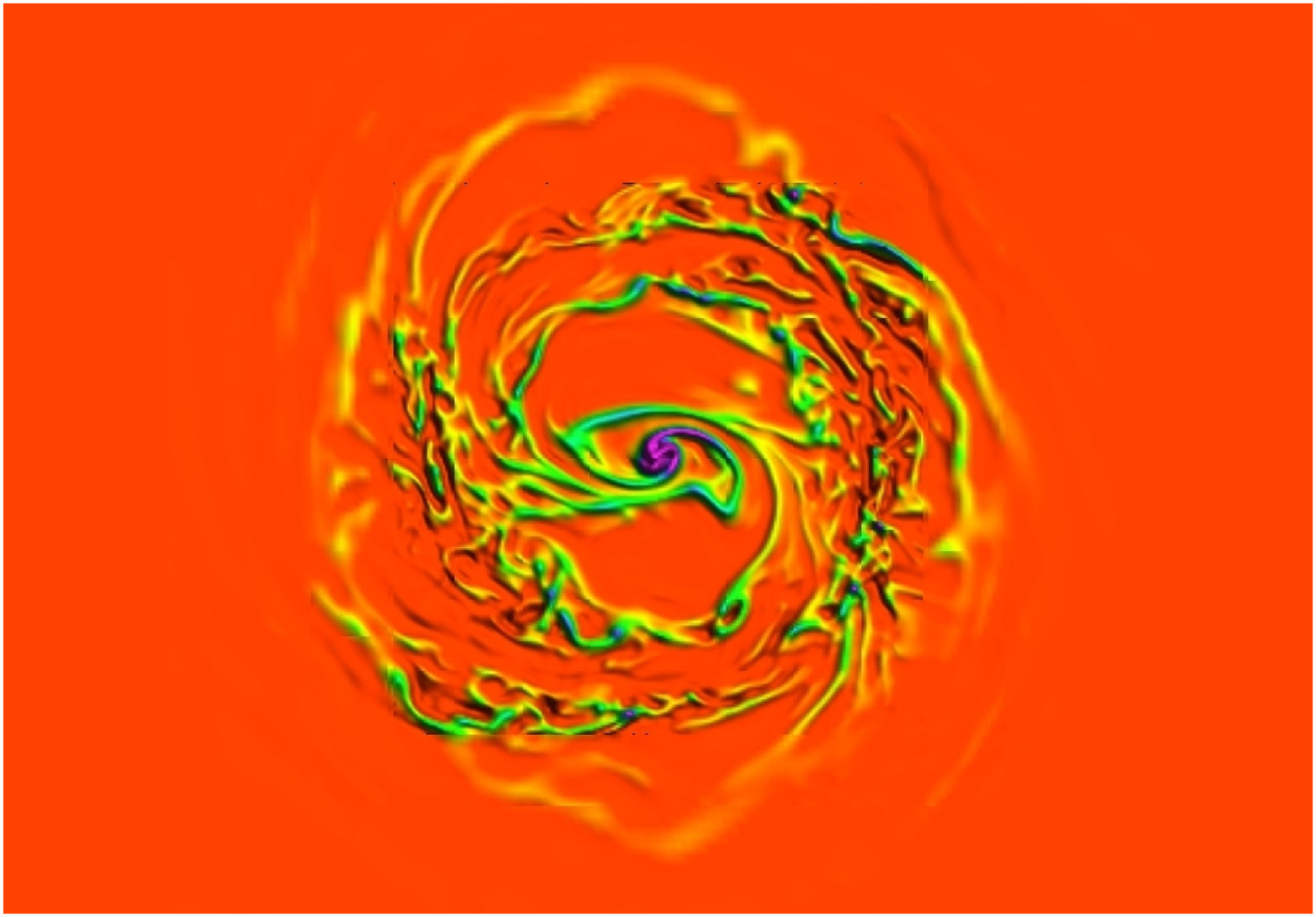}
\includegraphics[trim =5.5cm 2.20cm 5.5cm 2.20cm,height=1.4in,clip=true,draft=false]{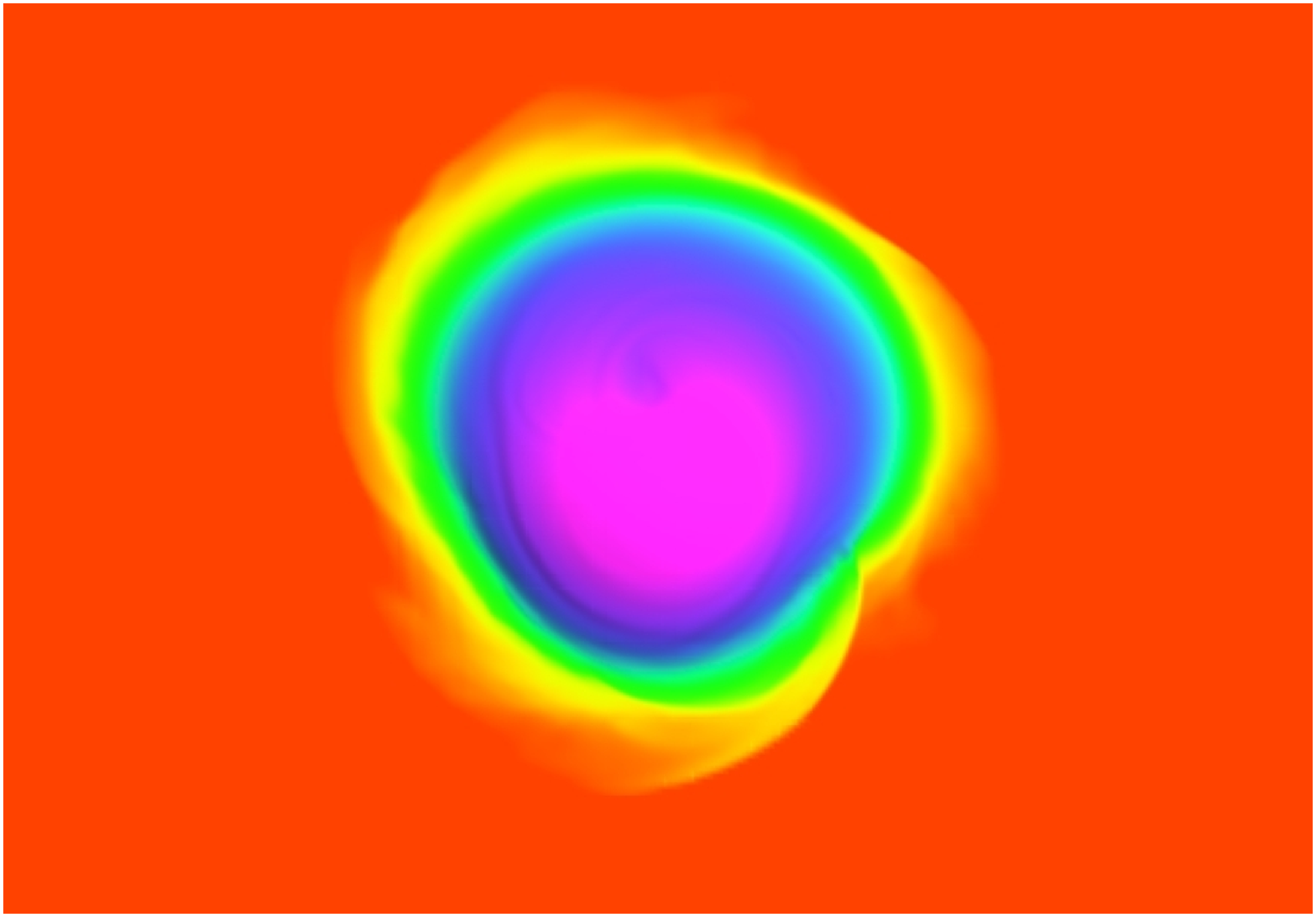}
\includegraphics[trim =5.5cm 2.20cm 5.5cm 2.20cm,height=1.4in,clip=true,draft=false]{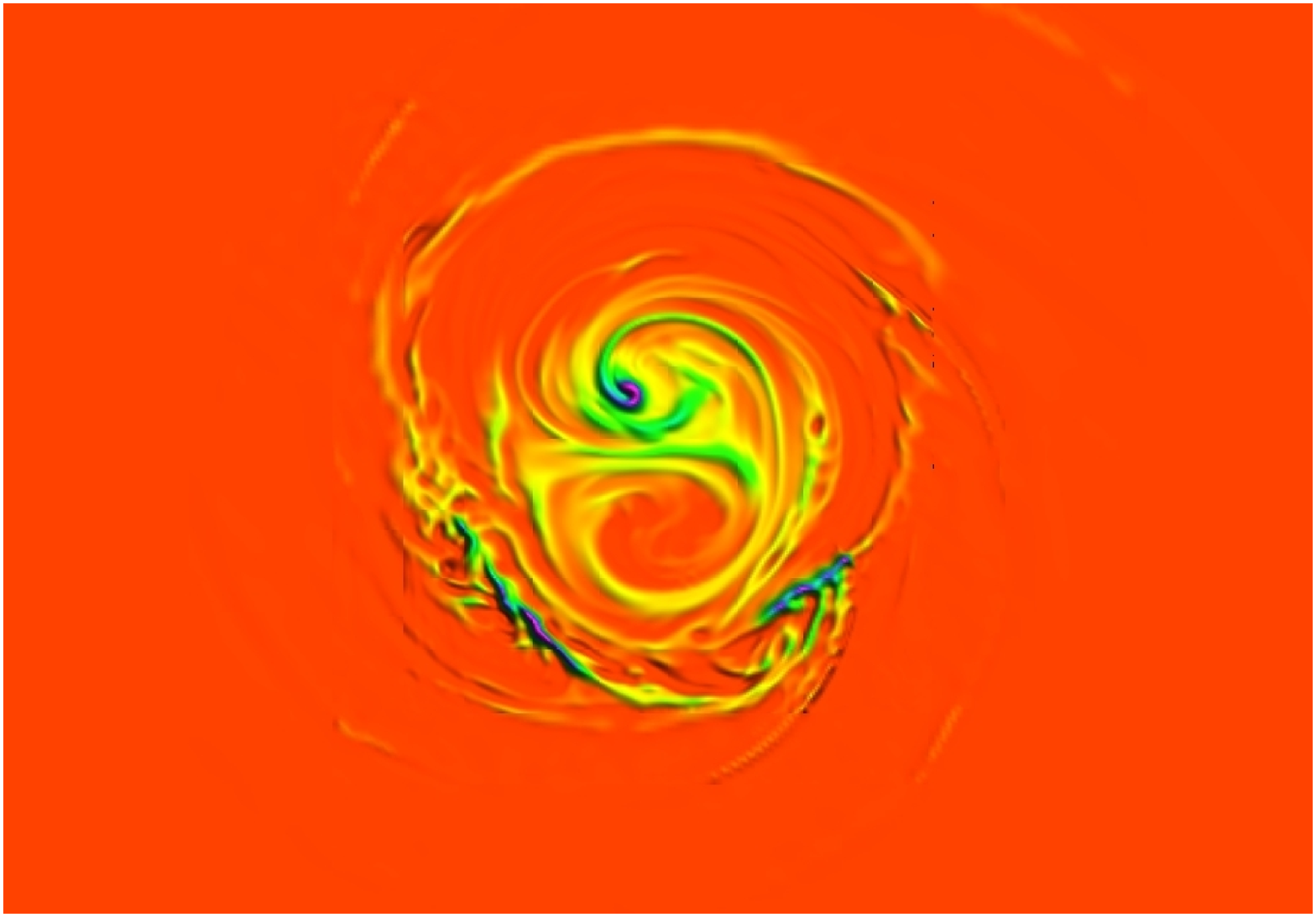}
\caption{Equatorial rest-mass density (left) and $\Omega_{xy}$ (right)
  snapshots at select times, advancing from top to bottom. At
  $t\approx3.1$ ms the NSs collide, leading to a vortex sheet. A
  couple of rotation periods later at $t\approx4.5$ ms two larger
  vortices form near the surface of the star at the shearing layers
  between the surface and the tidal tails.  These two vortices
  inspiral toward the center ($t\approx5.5$ ms) and merge
  ($t\approx6.5$ ms) creating an underdense center. This
  near-stationary, near-axisymmetric configuration persists for a few
  milliseconds, though the one-arm instability is now growing,
  eventually expelling the central vortex and associated underdensity
  from the center. By $t\approx14.6$ ms the instability is fully
  developed.  Each panel is $\approx50$ km per side.
  } \label{density_vorticity_snapshots}
\end{center}
\end{figure}


{\em Methodology.}---%
Our simulations are performed using the code described
in~\cite{code_paper}. The Einstein field equations are solved in the
generalized-harmonic formulation using finite difference methods,
while the matter is modeled as a perfect fluid with the corresponding
hydrodynamic equations evolved using
high-resolution shock-capturing techniques described
in~\cite{bhns_astro_paper}.

\begin{figure*}[t]
\begin{center} 
\includegraphics[trim =0.2cm 0.0cm 1.0cm 0.0cm,clip=true,height = 1.87in]{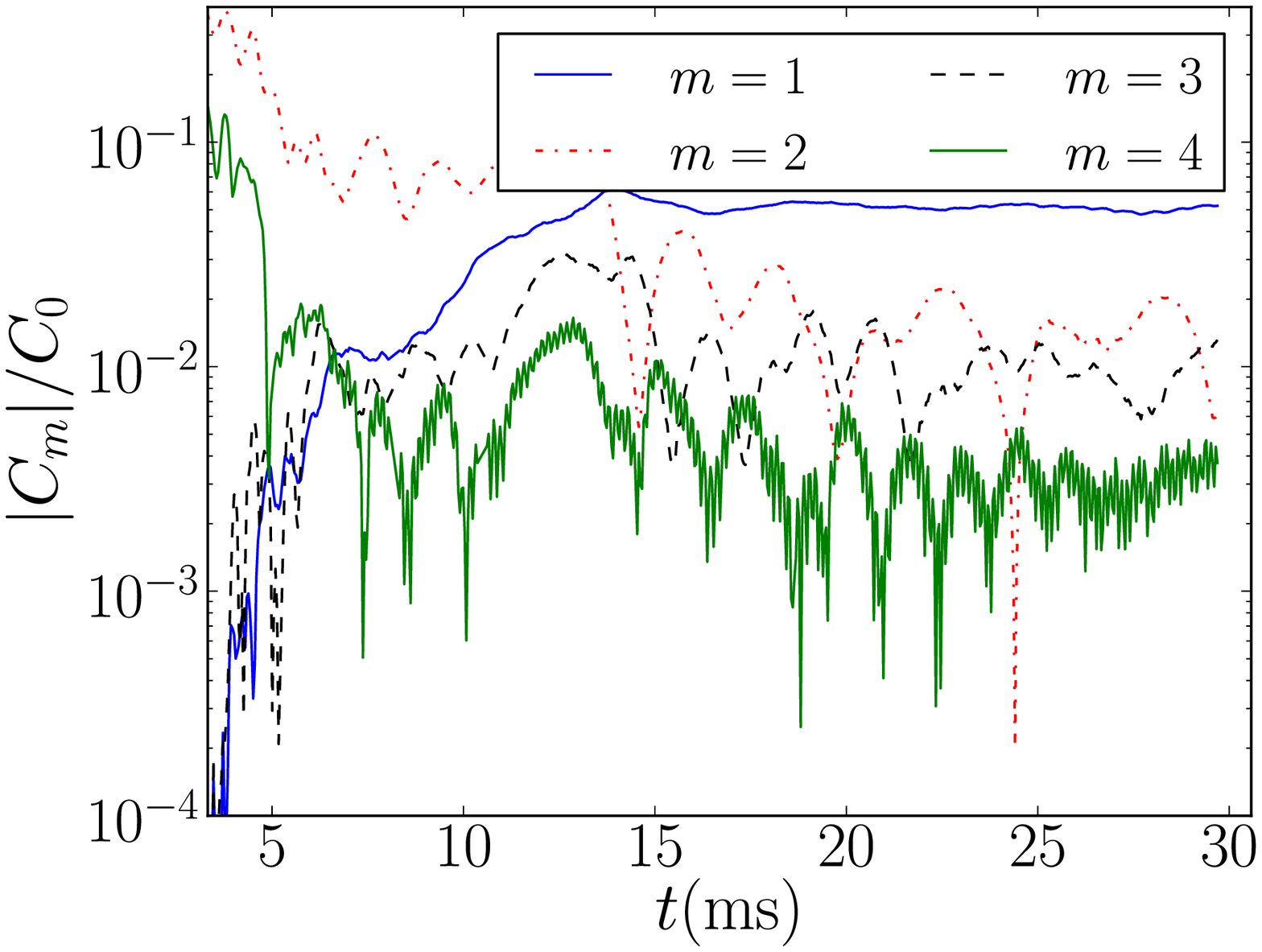}
\includegraphics[trim =2.0cm 0.cm 2.0cm 0.5cm,clip=true,height = 1.81in]{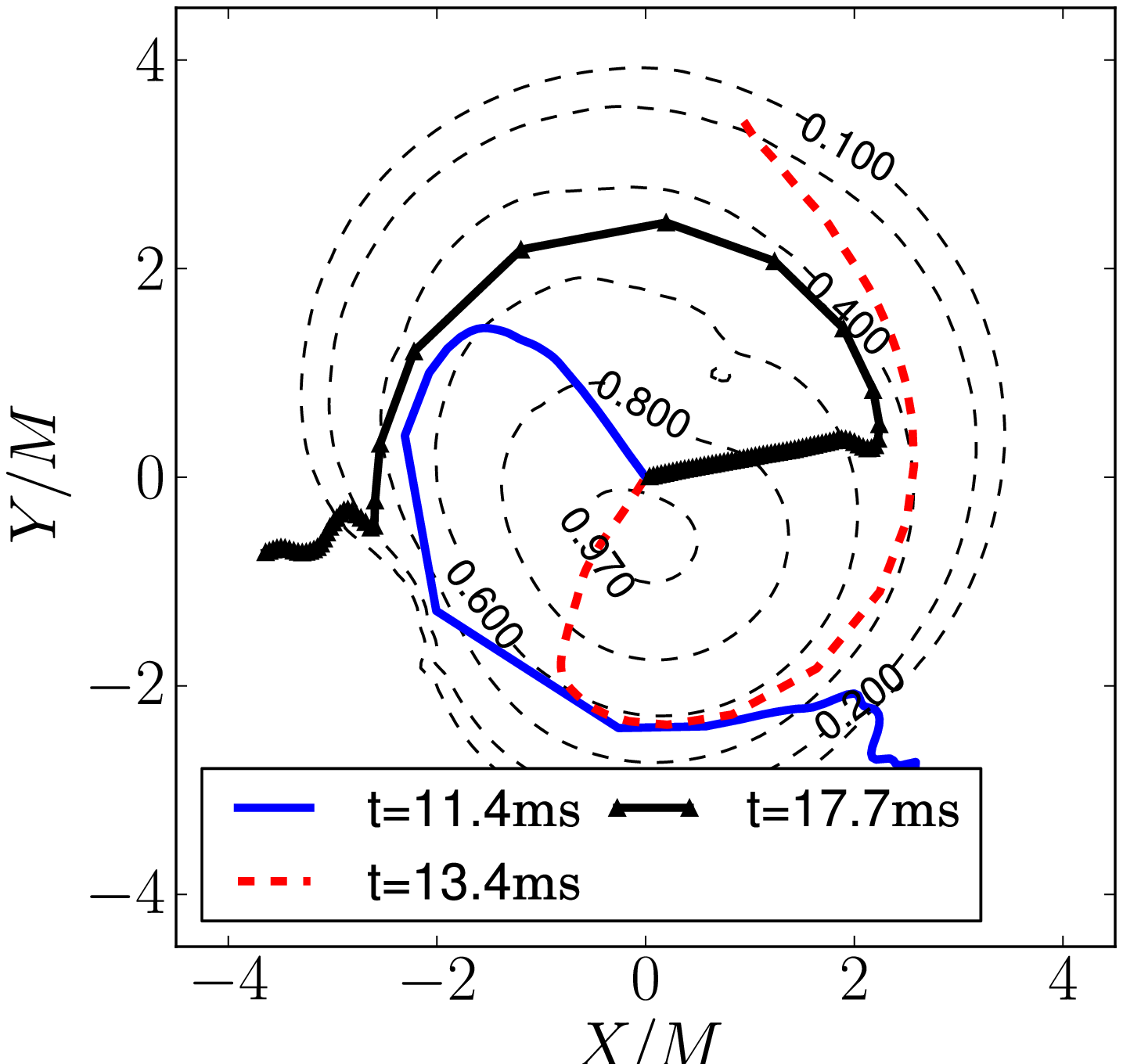}
\includegraphics[trim =0.4cm 0.0cm 1.0cm 0.0cm,clip=true,height = 1.87in]{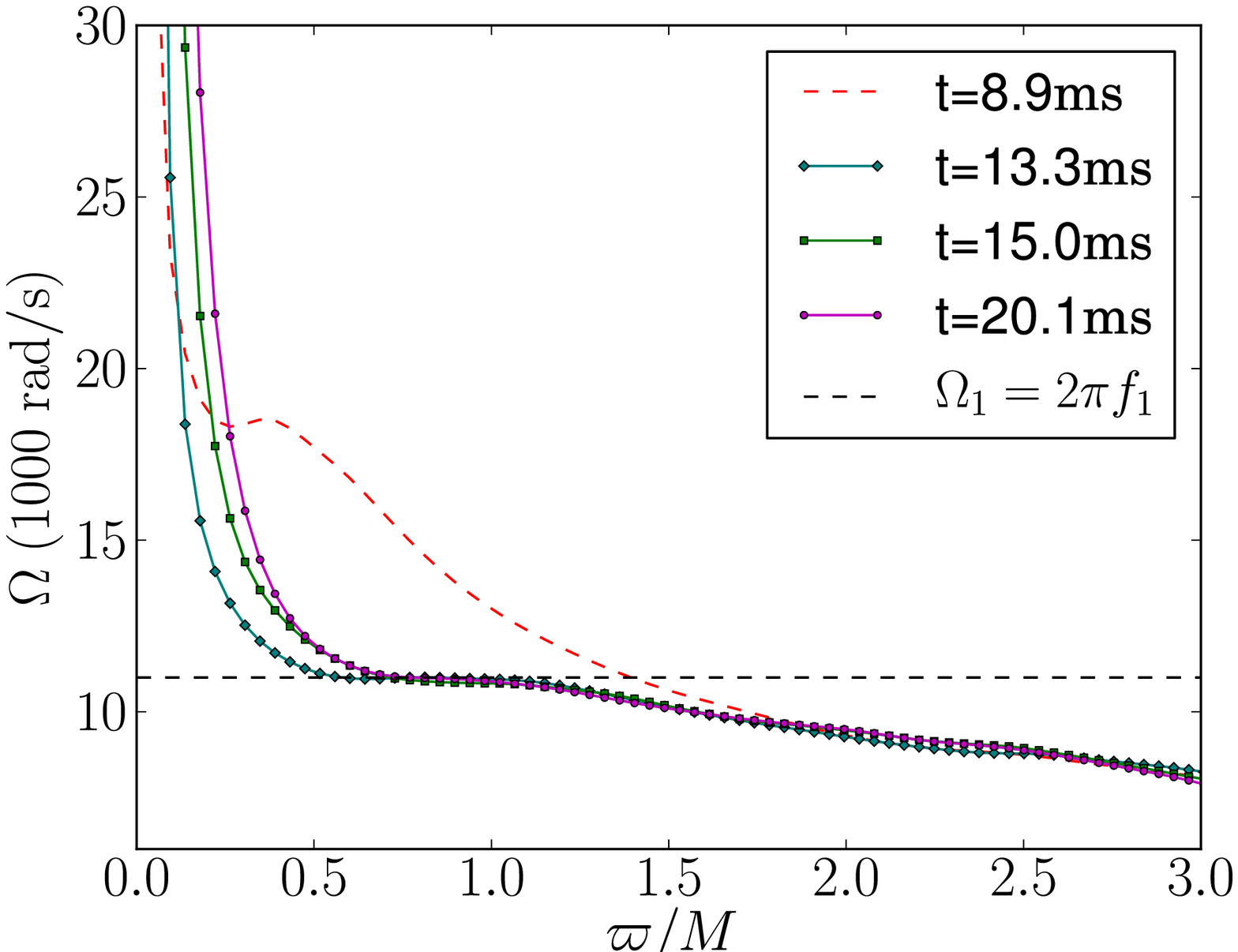}
\caption{Left: The magnitude of $C_m$ normalized to $C_0$.  Middle:
  The thick lines illustrate the phase of the mode $C_1$ as a function
  of radius $\varpi$ (specifically, we plot $X+iY=\varpi
  C_1(\varpi,0)/|C_1(\varpi,0)|$) at select times. Dashed thin lines
  are contours of $\rho_0$ at $t=13.4$ ms, normalized to its maximum
  value then.  The inlined numbers label the values of the level
  surfaces. The tiny contour at $X/M\approx Y/M \approx 1$ corresponds
  to a value of 0.6 and is at the location of the vortex. 
  Right: azimuthally averaged
  angular velocity at select times. Here $M=2.7M_\odot\simeq
  3.99$ km. Merger occurs at $t\simeq 3.0$ ms.
  } \label{C1modes}
\end{center} 
\end{figure*}

We prepare constraint-satisfying initial data as
in~\cite{idsolve_paper,East2012NSNS}, except that here the data
(matter and velocity profiles, and freely specifiable fields) are a
superposition of two rigidly-rotating, equilibrium NSs, generated with
the code of~\cite{1994ApJ...424..823C,1994ApJ...422..227C}. Each of
the stars has a mass of $1.35$ $M_\odot$ and dimensionless spin
$J_{\rm NS}/M_{\rm NS}^2=0.05$ (we adopt geometrized units with
$G=c=1$ throughout unless otherwise specified) aligned with the
ortital angular momentum. The initial separation is $d=50M\simeq200$
km [where $M$ is the total Arnowit-Deser-Misner (ADM) mass] and the
velocities and positions of the stars correspond to a marginally
unbound Newtonian orbit of periapse $r_p=8M$.
We adopt the ``HB" piece-wise polytropic cold EOS of~\cite{read} for
the matter, which yields a maximum static mass of 
$2.12$ $M_\odot$ ($2.53$ $M_\odot$ allowing for maximal uniform rotation).
For the evolution we incorporate a thermal component to the EOS,
$P_{\rm th}=0.5\epsilon_{\rm th}\rho_0$, to allow for
shock-heating. Here, $\epsilon_{\rm th}$ is the thermal component of
the specific internal energy $\epsilon$, and $\rho_0$ is the rest mass
density.

We employ adaptive mesh
refinement (AMR), where our hierarchy consists of six levels that are
dynamically adjusted during the evolution based on metric
truncation-error estimates. For convergence studies we perform the
simulations using three resolutions. All figures use data
from the highest resolution run, which has a base-level
grid of $321^3$ points and a finest level covering the eventual HMNS
diameter with $\sim 200$ points. The low and medium resolution runs
have $2$ and $1.5625\times$ the grid spacing, respectively.

To analyze the one-arm spiral instability we use several diagnostics.
We compute the azimuthal mode decomposition of the conserved
rest-mass density $C_m(\varpi,z) = \frac{1}{2\pi}\int_0^{2\pi}
\rho_0u^0\sqrt{-g} e^{im\phi}d\phi$, where $\varpi=\sqrt{x^2+y^2}$ is
the cylindrical coordinate radius, $\phi$ is the azimuthal angle,
$u^\mu$ is the fluid 4-velocity, and $g$ the determinant of the
spacetime metric. A similar quantity integrated throughout
the star is $C_m = \int \rho_0u^0\sqrt{-g} e^{im\phi}d^3x$. We 
follow the xy-component of the vorticity 2-form
$\Omega_{\mu\nu}=\nabla_\mu{(hu_\nu)}-\nabla_\nu{(hu_\mu)}$ on the
equatorial plane, where $\nabla_\mu$ is the covariant derivative, and
$h=1+\epsilon+P/\rho_0$ the specific enthalpy, with $P$ the
pressure. We also compute the ratios of total
kinetic ($T_{\rm kin}$)
and rotational kinetic energy ($T_{\rm rot}$)
to the gravitational potential energy $|W|$ as
in~\cite{1994ApJ...424..823C,1994ApJ...422..227C,BSNRbook,Kiuchi2012},
but in a coordinate center-of-mass frame: $x_{\rm
  cm}^i=\frac{1}{C_0}\int x^i\rho_0u^0\sqrt{-g}d^3x$.  These
diagnostics are not gauge independent, but are nevertheless useful in
identifying qualitative features of the one-arm spiral instability.


{\em Results.}---%
Following merger we find a long-lived HMNS that is subject to the
one-armed spiral instability.  In
Fig.~\ref{density_vorticity_snapshots} we show equatorial $\rho_0$ and
$\Omega_{xy}$ snapshots illustrating the dynamics.  Two larger
vortices form near the surface of the HMNS from shearing with the
tidal tails, and subsequently spiral towards the center and merge,
creating an underdensity around the rotation axis.  (Numerous other
smaller vortices also form during the early stage of the merger, in
particular interior to the HMNS following break-up of the vortex sheet
formed at first contact, but for the most part they are quickly
stretched away and do not seem to play a significant role in creating
the underdense core.)
The one-arm spiral instability is triggered around this time, in
agreement with earlier Newtonian simulations~\cite{Saijo2003} which
suggest that such a toroidal HMNS configuration is a necessary
condition for the instability.  This is consistent with the growth of
the $C_1$ mode shown in the left panel of Fig.~\ref{C1modes}. Several
milliseconds after formation of the underdense core, $\sim11$ ms after
merger, $C_1$ has grown to saturation, dominating the azimuthal modes
of the HMNS.
%
%
%
%
We can characterize the approximate growth rate of the
instability by noting that it takes $\approx 1.2$ ms for this mode to
grow from $1/4$ to $1/2$ its saturation level. From the Fourier
transform of $C_m$ we determine the dominant frequencies $f_m$ of the
density modes to be $f_1=1.75$ kHz, $f_2=3.4$ kHz, and $f_3=5.2$ kHz,
i.e., $f_m\approx m\times f_1$. The time to saturation and frequency
of the $m=1$ mode differ by at most $15\%$ and $2\%$ among the different
resolutions, respectively. 

The characteristic one-arm spiral pattern of the instability can be seen in the middle
panel of Fig.~\ref{C1modes}, which shows the phase of the $m=1$ mode
in the equatorial plain. 
The right panel in Fig.~\ref{C1modes} plots
the azimuthally averaged angular velocity 
of the fluid in the HMNS as a function of radius at several times.
If we take the angular frequency $2\pi f_1$ (horizontal line in the
panel) calculated above from the time dependence of $C_1$ to be the
oscillation frequency of the unstable mode, the right panel shows that
there exists a corotation radius at $\varpi\approx 1.4$M prior to the
development of the instability. Following saturation of the
instability, the region $0.5M \lesssim \varpi\lesssim 1.2M$ rotates
almost rigidly with this same angular frequency.

After the HMNS settles from the violence of the merger (by $t\simeq 7$
ms),
we find $T_{\rm kin}/|W|\approx T_{\rm rot}/|W|\approx 0.26$, and steadily drops to
$0.23$ as the instability saturates. Thus, to within gauge ambiguities, the
instability cannot be classified as a low-$T/|W|$ instability, but $T/|W|$ is
slightly below the usual threshold for the dynamical bar mode
instability~\cite{DynamicalBarmodeOrig}.

\begin{figure} 
\begin{center} 
\includegraphics[clip=true,width=3.2in]{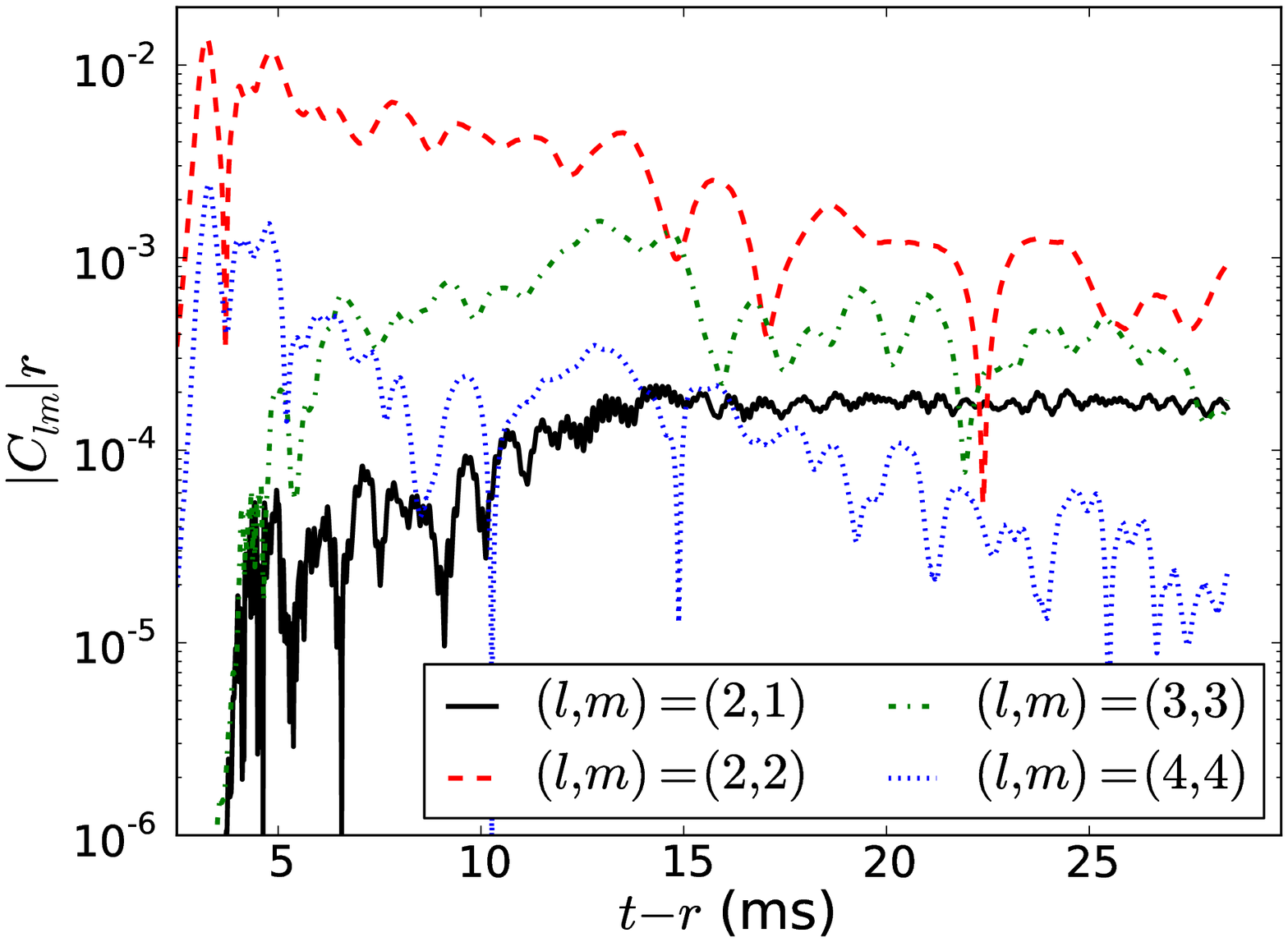} 
\includegraphics[clip=true,width=3.2in]{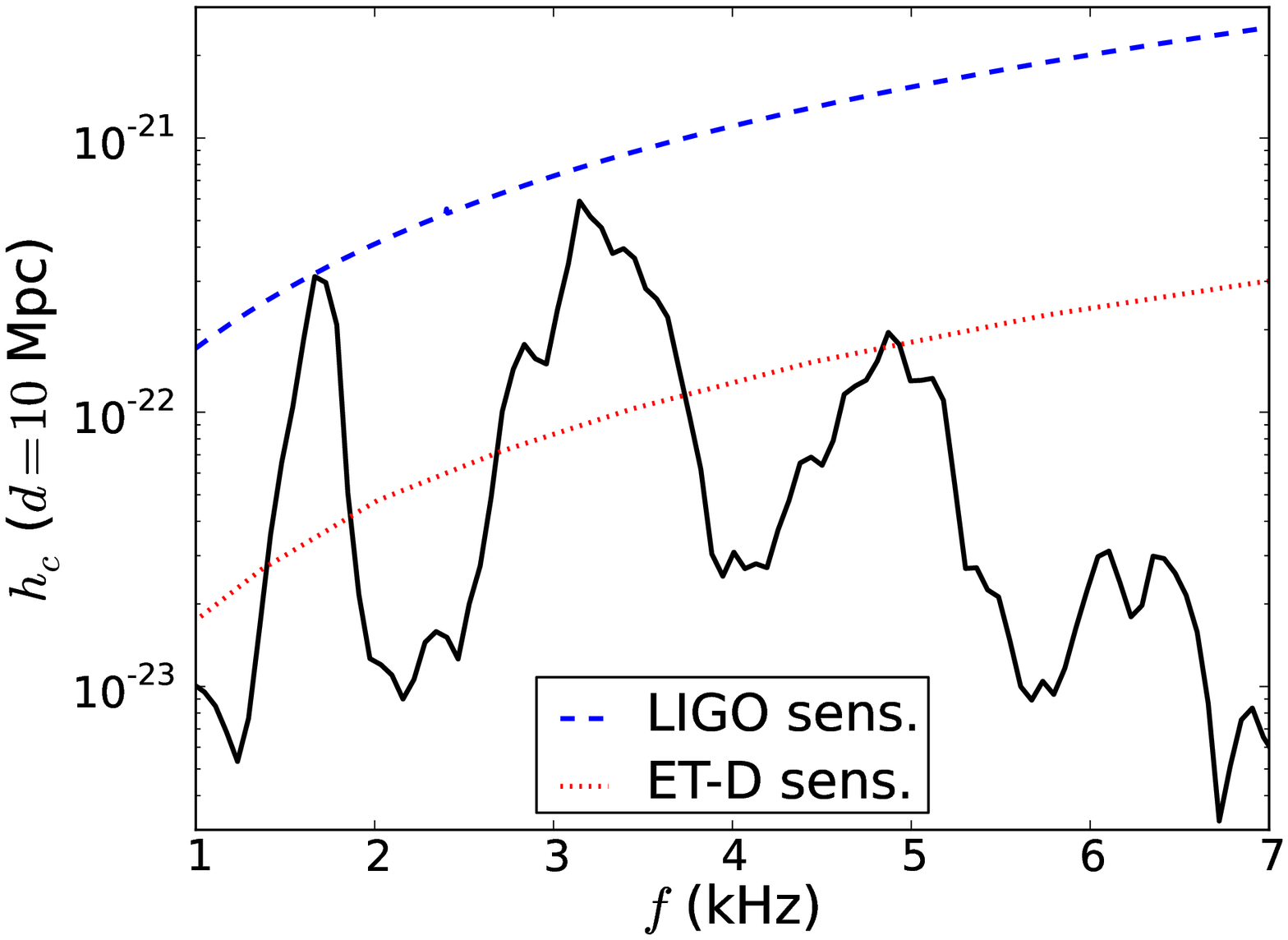}
\caption{ Top: the amplitude of spin-weight -2 spherical harmonic components of
the post-merger GW signal. Bottom: the full GW spectrum from the last $\sim 15$
ms  (when the instability is fully developed), as would be seen
by an edge-on observer $10$ Mpc away. Also plotted are the aLIGO and proposed
Einstein Telescope (ET-D) sensitivity curves~\cite{ET_D}.  If the $m=1$ mode
persists as long as the HMNS lifetime $t_{\rm HMNS}$, estimated to be $\sim
O(1)$ s, the peak power at $\sim 1.7$ kHz could be enhanced by a factor $t_{\rm
HMNS}/(15 \ {\rm ms})\sim O(10^2)$.
\label{gw_spec_post_merger} }
\end{center} 
\end{figure}

The upper panel of Fig.~\ref{gw_spec_post_merger} shows the leading
spherical harmonic components of the post-merger GW signal. The
$(\ell,m)=(2,1)$ mode mirrors the growth and saturation of the $C_1$
density perturbation,
but remains sub-dominant compared to the $(\ell,m)=(2,2)$ and $(\ell,m)=(3,3)$
over the time simulated. However, in terms of detectability as the GW power
spectrum shown in the lower panel of Fig.~\ref{gw_spec_post_merger} indicates,
the less GW power in the $m=1$ mode is in part offset by the lower frequency of
the mode 
where ground-based detectors have greater sensitivity.

Comparison with the aLIGO sensitivity curve in Fig.~\ref{gw_spec_post_merger}
shows the part of the GW signal due to the one-arm instability is too weak for
likely detection, unless the HMNS and excited $m=1$ mode can persist for a time
considerably longer than the length of the simulation.
From Figs.~\ref{C1modes} and~\ref{gw_spec_post_merger} we note that
after saturation the $m=1$ mode and corresponding GW signal persists
at roughly constant amplitude till the end of the simulation, in
contrast to the other modes that trend to decay.  Thus, the $m=1$
component of the signal may last much longer than the $\approx 15$ ms
of integration used in Figs.~\ref{gw_spec_post_merger}, possibly even
the entire lifetime of the HMNS before collapse to a BH. A rough
estimate of this lifetime calculated from the rate of angular momentum
loss to GWs and amount of unradiated angular momentum at the end of
the simulation gives $t_{\rm HMNS}=J_{\rm ADM}/\dot J_{\rm GW} \sim 2$
s, which could give an additional factor of $O(10^2)$ in GW power and
make the mode detectable by aLIGO out to $\approx 10$ Mpc and by the
ET out to  $\approx 100$ Mpc.

{\em Concluding Remarks.}---%
%
%
%
%
First, some caveats related to the numerics are in order. Though we do
see the expected second-order convergence for the pre-merger epoch,
our resolution sequence is not high enough to show the expected
first-order (due to shocks) convergence for the post-merger epoch in
certain quantities.  This is likely because with higher resolution we
observe the appearance of ever smaller scale vortices following
merger, and it is very challenging to achieve convergence in such a
turbulent-like environment.  However, essential qualitative features of
the post-merger remnant appear robust, most importantly, the
development of the one-arm instability and its order-of-magnitude
growth time, frequency and saturation amplitude.  On the other hand,
our low resolution run forms a BH $\sim 19$ ms following merger,
whereas the medium and high resolution have not, even after the $\sim
27$ ms they were continued post merger. This suggests our order of
magnitude estimate above for the lifetime of the HMNS may be too
optimistic.  However, there are many factors that will affect the
actual lifetime of a HMNS, including physical effects we do not model
(e.g. magnetic fields and neutrino cooling), parameters quantifying
uncertainty in the EOS (for stiff EOSs typical HMNS remnants may
possibly survive for $\sim 2-3$ s~\cite{PhysRevLett.107.051102}), and
the broader range of relevant initial conditions (e.g. mass ratio,
lower eccentricity at merger, spin orientation). Conversely, strong
sensitivity of the lifetime of the HMNS to properties of the system
means greater possibility of measuring related quantities from
putative future multi-messenger observations.


More details on convergence, other cases, and other properties of
these mergers
will be presented in~\cite{followup}. Important questions for future
work are what elements of the particular case studied here are
essential to give rise to the instability, and why it was not present
and/or pointed out in previous studies.
It could be that only some combination of orbital eccentricities, a
particular EOS, component masses and spins lead to a long-lived,
unstable HMNS. Alternatively, these factors could affect the growth
rate such that the structure of the instability was not clearly seen
by the termination of earlier simulations.  For example,
~\cite{Bernuzzi:2013rza} reported strong $m=1$ modes in HMNS remnants
from quasicircular NSNS mergers with spinning NSs (employing a
$\Gamma$-Law EOS and initial $1.5$ $M_\odot$ NSs), though there they
were ascribed as likely due to ``mode couplings''.  The $t>100$ ms
post-merger evolution of a HMNS presented
in~\cite{Rezzolla2010CQGra..27k4105R} would not have seen any odd-$m$
instabilities due to the $\pi$ symmetry enforced there.
It is important to resolve these issues for quasi-circular mergers
involving spinning NSs, for as discussed in the introduction, these
are the most likely sources of observable GWs from dynamically
assembled systems in GCs.  We plan to address many of these issues in
future work.

\acknowledgments

It is a pleasure to thank Tom Abel, Andreas Bauswein, and Roman Gold
for useful discussions.  This work was supported by the Simons
Foundation and NSF grant PHY-1305682 at Princeton University, as well
as NSF grant PHY-1300903 and NASA grant NNX13AH44G at the University
of Illinois at Urbana-Champaign. Computational resources were provided
by XSEDE/TACC under grants TG-PHY100053, TG-MCA99S008, and the Orbital
cluster at Princeton University.

\bibliographystyle{h-physrev}
\bibliography{ref}

\end{document}